\documentclass[aps,pre,showpacs,showkeys,onecolumn,superscriptaddress,amsmath,amssymb]{revtex4}
\usepackage[dvips]{graphicx}
\usepackage[dvips]{color}

\newcommand{\ie}{\emph{i.e.}, }
\newcommand{\TT}{\mathrm{T}}
\newcommand{\Tr}{\mathrm{Tr}}
\newcommand{\Det}{\mathrm{Det}}
\newcommand{\dd}{\mathrm{d}}
\newcommand{\DD}{\mathrm{D}}

\newcommand{\im}{\mathrm{Im}}
\newcommand{\ii}{\mathrm{i}}
\newcommand{\ee}{\mathrm{e}}
\newcommand{\diag}{\mathrm{diag}}
\newcommand{\la}{\left\langle}
\newcommand{\ra}{\right\rangle}
\newcommand{\MAq}{\mathrm{MA} ( q )}
\newcommand{\VMAq}{\mathrm{VMA} ( q )}
\newcommand{\VMAqOne}{\mathrm{VMA} ( q_{1} )}
\newcommand{\VMAqTwo}{\mathrm{VMA} ( q_{2} )}
\newcommand{\VMAOne}{\mathrm{VMA} ( 1 )}
\newcommand{\ARq}{\mathrm{AR} ( q )}
\newcommand{\VARq}{\mathrm{VAR} ( q )}
\newcommand{\VARqOne}{\mathrm{VAR} ( q_{1} )}
\newcommand{\VAROne}{\mathrm{VAR} ( 1 )}
\newcommand{\VARMA}{\mathrm{VARMA} ( q_{1} , q_{2} )}
\newcommand{\VARMAOneOne}{\mathrm{VARMA} ( 1 , 1 )}

\begin{document}

\title{A Random Matrix Approach to VARMA Processes}
\author{Zdzis\l{}aw Burda}
\email{zdzislaw.burda@uj.edu.pl}
\affiliation{Marian Smoluchowski Institute of Physics and Mark Kac Complex Systems Research Centre, Jagiellonian University, Reymonta 4, 30--059 Krak\'{o}w, Poland}
\author{Andrzej Jarosz}
\email{andrzej.jarosz@clico.pl}
\affiliation{Clico Ltd., Oleandry 2, 30--063 Krak\'{o}w, Poland}
\author{Maciej A. Nowak}
\email{nowak@th.if.uj.edu.pl}
\affiliation{Marian Smoluchowski Institute of Physics and Mark Kac Complex Systems Research Centre, Jagiellonian University, Reymonta 4, 30--059 Krak\'{o}w, Poland}
\author{Ma\l{}gorzata Snarska}
\email{snarska@th.if.uj.edu.pl}
\affiliation{Marian Smoluchowski Institute of Physics and Mark Kac Complex Systems Research Centre, Jagiellonian University, Reymonta 4, 30--059 Krak\'{o}w, Poland}
\affiliation{Cracow University of Economics, Department of Econometrics and Operations Research, Rakowicka 27, 31--510 Krak\'{o}w, Poland}

\date{\today}

\setlength{\parindent}{4ex}
\setlength{\parskip}{1.5ex plus 0.5ex minus 0.2ex}
\topmargin=0cm

\begin{abstract}
We apply random matrix theory to derive spectral density of large sample covariance matrices generated by multivariate $\VMAq$, $\VARq$ and $\VARMA$ processes. In particular, we consider a limit where the number of random variables $N$ and the number of consecutive time measurements $T$ are large but the ratio $N / T$ is fixed. In this regime the underlying random matrices are asymptotically equivalent to Free Random Variables (FRV). We apply the FRV calculus to calculate the eigenvalue density of the sample covariance for several VARMA--type processes. We explicitly solve the $\VARMAOneOne$ case and demonstrate a perfect agreement between the analytical result and the spectra obtained by Monte Carlo simulations. The proposed method is purely algebraic and can be easily generalized to \smash{$q_{1} > 1$} and \smash{$q_{2} > 1$}.
\end{abstract}


\maketitle


\section{Introduction}
\label{s:Introduction}

Vector auto--regressive (VAR) models play an important role in contemporary macro--economics, being an example of an approach called the ``dynamic stochastic general equilibrium'' (DSGE), which is superseding traditional large--scale macro--econometric forecasting methodologies~\cite{Sims1980}. The motivation behind them is based on the assertion that more recent values of a variable are more likely to contain useful information about its future movements than the older ones. On the other hand, a standard tool in multivariate time series analysis is vector moving average (VMA) models, which is really a linear regression of the present value of the time series w.r.t. the past values of a white noise. A broader class of stochastic processes used in macro--economy comprises both these kinds together in the form of vector auto--regressive moving average (VARMA) models. These methodologies can capture certain spatial and temporal structures of multidimensional variables which are often neglected in practice; including them not only results in more accurate estimation, but also leads to models which are more interpretable. They are widely used by academia and central banks (\emph{cf.} the European Central Bank's Smets--Wouters model for the euro zone~\cite{SmetsWouters2002}), as they constitute quite a simple version of the DSGE equations.

VARMA models are constructed from a number of univariate ARMA (Box--Jenkins; see for example~\cite{BoxJenkinsReinsel1994}) processes, typically coupled with each other. In this paper, we investigate only a significantly simplified circumstance when there is no coupling between the many ARMA components. One may argue that this is too far fetched and will be of no use in describing an economic reality. However, one may also treat it as a ``zeroth--order hypothesis,'' analogously to the idea of~\cite{LalouxCizeauBouchaudPotters1999,PlerouGopikrishnanRosenowAmaralStanley1999} in finance, namely that the case with no cross--covariances is considered theoretically, and subsequently compared to some real--world data modeled by a VARMA process; any discrepancy between the two will reflect nontrivial cross--covariances present in the system, thus permitting their investigation. This latter route is taken in this communication.

A challenging and yet increasingly important problem is the estimation of large covariance matrices generated by these stationary \smash{$\VARMA$} processes, since high dimensionality of the data as compared to the sample size is quite common in many statistical problems (the ``dimensionality curse''). Therefore, an appropriate ``noise cleaning'' procedure has to be implemented, and random matrix theory (RMT) provides a natural and efficient outfit for doing that. In particular, the mean spectral densities (a.k.a. ``limiting spectral distributions,'' LSD) of the Pearson estimators of the cross--covariances for the $\VMAOne$ and $\VAROne$ models, in the relevant high--dimensionality sector and under the full decoupling, have been derived in~\cite{JinWangMiaoHuang2009} by applying the framework proposed by~\cite{BaiSilverstein2006}.

In this paper, we suggest that such calculations can be considerably simplified by resorting to a mathematical concept of the free random variables (FRV) calculus~\cite{VoiculescuDykemaNica1992,Speicher1994}, succinctly introduced in sec.~\ref{s:DoublyCorrelatedWishartEnsemblesAndFreeRandomVariables}. Our general FRV formula~\cite{BurdaJaroszJurkiewiczNowakPappZahed2009} allows not only to rediscover, which much less strain, the two fourth--order polynomial equations obtained in~\cite{JinWangMiaoHuang2009} in the $\VMAOne$ and $\VAROne$ cases, but also to derive a sixth--order equation (\ref{eq:VARMAOneOneMainEquation}) which produces the mean spectral density for a more involved $\VARMAOneOne$ model. The results are verified by numerical simulations, which show a perfect agreement. This is all done in sec.~\ref{s:VARMAFromFreeRandomVariables}.


\section{Doubly Correlated Wishart Ensembles and Free Random Variables}
\label{s:DoublyCorrelatedWishartEnsemblesAndFreeRandomVariables}


\subsection{Doubly Correlated Wishart Ensembles}
\label{ss:DoublyCorrelatedWishartEnsembles}


\subsubsection{Correlated Gaussian Random Variables}
\label{sss:CorrelatedGaussianRandomVariables}

$\VARMA$ stochastic processes, as we will see below, fall within quite a general set--up encountered in many areas of science where a probabilistic nature of multiple degrees of freedom evolving in time is relevant, for example, multivariate time series analysis in finance, applied macro--econometrics and engineering. To describe this framework, consider a situation of $N$ time--dependent random variables which are measured at $T$ consecutive time moments (separated by some time interval $\delta t$); let \smash{$Y_{i a}$} be the value od the $i$--th ($i = 1 , \ldots , N$) random number at the $a$--th time moment ($a = 1 , \ldots , T$); together, they make up a rectangular $N \times T$ matrix $\mathbf{Y}$. In what usually would be the first approximation, each \smash{$Y_{i a}$} is supposed to be drawn from a Gaussian probability distribution. We will also assume that they have mean values zero, \smash{$\langle Y_{i a} \rangle = 0$}. These degrees of freedom may in principle display mutual correlations. A set of correlated zero--mean Gaussian numbers is fully characterized by the two--point covariance function, \smash{$\mathcal{C}_{i a , j b} \equiv \langle Y_{i a} Y_{j b} \rangle$} if the underlying stochastic process generating these numbers is stationary. Linear stochastic processes, including $\VARMA$, belong to this category. We will restrict our attention to an even narrower class where the cross--correlations between different variables and the auto--correlations between different time moments are factorized, \ie
\begin{equation}\label{eq:CovarianceFunctionFactorization}
\la Y_{i a} Y_{j b} \ra = C_{i j} A_{a b} .
\end{equation}
In this setting, the inter--variable covariances do not change in time (and are described by an $N \times N$ cross--covariance matrix $\mathbf{C}$), and also the temporal covariances are identical for all the numbers (and are included in a $T \times T$ auto--covariance matrix $\mathbf{A}$; both these matrices are symmetric and positive--definite). The Gaussian probability measure with this structure of covariances is known from textbooks,
$$
P_{\mathrm{c.G.}} ( \mathbf{Y} ) \DD \mathbf{Y} = \frac{1}{\mathcal{N}_{\mathrm{c.G.}}} \exp \left( - \frac{1}{2} \sum_{i , j = 1}^{N} \sum_{a , b = 1}^{T} Y_{i a} \left[ \mathbf{C}^{- 1} \right]_{i j} Y_{j b} \left[ \mathbf{A}^{- 1} \right]_{b a} \right) \DD \mathbf{Y} =
$$
\begin{equation}\label{eq:CorrelatedGaussian}
= \frac{1}{\mathcal{N}_{\mathrm{c.G.}}} \exp \left( - \frac{1}{2} \Tr \mathbf{Y}^{\TT} \mathbf{C}^{- 1} \mathbf{Y} \mathbf{A}^{- 1} \right) \DD \mathbf{Y} ,
\end{equation}
where the normalization constant \smash{$\mathcal{N}_{\mathrm{c.G.}} = ( 2 \pi )^{N T / 2} ( \Det \mathbf{C} )^{T / 2} ( \Det \mathbf{A} )^{N / 2}$}, and the integration measure \smash{$\DD \mathbf{Y} \equiv \prod_{i = 1}^{N} \prod_{a = 1}^{T} \dd Y_{i a}$}, while the letters ``c.G.'' stand for ``correlated Gaussian.''

Now, a standard way to approach correlated Gaussian random numbers is to recall that they can always be decomposed as linear combinations of uncorrelated Gaussian degrees of freedom; indeed, this is achieved through the transformation
\begin{equation}\label{eq:UncorrelatedGaussian}
\mathbf{Y} = \sqrt{\mathbf{C}} \widetilde{\mathbf{Y}} \sqrt{\mathbf{A}} , \qquad \textrm{which yields} \qquad P_{\mathrm{G.}} ( \widetilde{\mathbf{Y}} ) \DD \widetilde{\mathbf{Y}} = \frac{1}{\mathcal{N}_{\mathrm{G.}}} \exp \left( - \frac{1}{2} \Tr \widetilde{\mathbf{Y}}^{\TT} \widetilde{\mathbf{Y}} \right) \DD \widetilde{\mathbf{Y}} ,
\end{equation}
where the square roots of the covariance matrices, necessary to facilitate the transition, exist due to the positive--definiteness of $\mathbf{C}$ and $\mathbf{A}$; the new normalization reads \smash{$\mathcal{N}_{\mathrm{G.}} = ( 2 \pi )^{N T / 2}$}.


\subsubsection{Estimating Equal--Time Cross--Covariances}
\label{sss:EstimatingEqualTimeCrossCovariances}

An essential problem in multivariate analysis is to determine (estimate) the covariance matrices $\mathbf{C}$ and $\mathbf{A}$ from given $N$ time series of length $T$ of the realizations of our random variables \smash{$Y_{i a}$}. For simplicity, we do not distinguish in notation between random numbers, \ie the population, and their realizations in actual experiments, \ie the sample. Since the realized cross--covariance between degrees $i$ and $j$ at the same time $a$ is \smash{$Y_{i a} Y_{j a}$}, the simplest method to estimate the today's cross--covariance \smash{$c_{i j}$} is to compute the time average,
\begin{equation}\label{eq:EstimatorcDefinition}
c_{i j} \equiv \frac{1}{T} \sum_{a = 1}^{T} Y_{i a} Y_{j a} , \qquad \textrm{\ie} \qquad \mathbf{c} = \frac{1}{T} \mathbf{Y} \mathbf{Y}^{\TT} = \frac{1}{T} \sqrt{\mathbf{C}} \widetilde{\mathbf{Y}} \mathbf{A} \widetilde{\mathbf{Y}}^{\TT} \sqrt{\mathbf{C}} .
\end{equation}
This is usually named the ``Pearson estimator'', up to the prefactor which depending on the context is $1 / ( T - 1 )$ or $1 / T$. Other estimators might be introduced, such as between distinct degrees of freedom at separate time moments (``time--delayed estimators''), or with certain decreasing weights given to older measurements to reflect their growing obsolescence (``weighted estimators''), but we will not investigate them in this article. Furthermore, in the last equality in (\ref{eq:EstimatorcDefinition}), we cast $\mathbf{c}$ through the uncorrelated Gaussian numbers contained in \smash{$\widetilde{\mathbf{Y}}$}, the price to pay for this being that the covariance matrices now enter into the expression for $\mathbf{c}$, making it more complicated; this will be the form used hereafter. The random matrix $\mathbf{c}$ is called a ``doubly correlated Wishart ensemble''~\cite{Wishart1928}.

Let us also mention that the auto--covariance matrix $\mathbf{A}$ can be estimated through \smash{$\mathbf{a} \equiv ( 1 / N ) \mathbf{Y}^{\TT} \mathbf{Y}$}. However, it is verified that this object carries identical information to the one contained in $\mathbf{c}$ (it is ``dual'' to $\mathbf{c}$), and therefore may safely be discarded. Indeed, these two estimators have same non--zero eigenvalues (modulo an overall rescaling by $r$), and the larger one has $| T - N |$ additional zero modes.

Any historical estimator is inevitably marred by the measurement noise; it will reflect the true covariances only to a certain degree, with a superimposed broadening due to the finiteness of the time series. More precisely, there are $N ( N + 1 ) / 2$ independent elements in $\mathbf{C}$, to be estimated from $N T$ measured quantities $\mathbf{Y}$, hence the estimation accuracy will depend on the ``rectangularity ratio,''
\begin{equation}\label{eq:NoiseToSignalRatio}
r \equiv \frac{N}{T} ;
\end{equation}
the closer $r$ to zero, the more truthful the estimate. This is a cornerstone of classical multivariate analysis. Unfortunately, a practical situation will typically feature a large number of variables sampled over a comparably big number of time snapshots, so that we may approximately talk about the ``thermodynamical limit,''
\begin{equation}\label{eq:ThermodynamicalLimit}
N \to \infty , \qquad T \to \infty , \qquad \textrm{such that} \qquad r = \textrm{fixed} .
\end{equation}
On the other hand, it is exactly this limit in which the FRV calculus (see the subsection below for its brief elucidation) can be applied; hence, the challenge of de--noising is somewhat counterbalanced by the computationally powerful FRV techniques.


\subsection{A Short Introduction to the Free Random Variables Calculus: The Multiplication Algorithm}
\label{ss:AShortIntroductionToTheFreeRandomVariablesCalculusTheMultiplicationAlgorithm}


\subsubsection{The $M$--Transform and the Spectral Density}
\label{sss:TheMTransformAndTheSpectralDensity}

Any study of a (real symmetric $K \times K$) random matrix $\mathbf{H}$ will most surely include a fundamental question about the average values of its (real) eigenvalues \smash{$\lambda_{1} , \ldots , \lambda_{K}$}. They are concisely encoded in the ``mean spectral density,''
\begin{equation}\label{eq:MeanSpectralDensityDefinition}
\rho_{\mathbf{H}} ( \lambda ) \equiv \frac{1}{K} \sum_{i = 1}^{K} \la \delta \left( \lambda - \lambda_{i} \right) \ra = \frac{1}{K} \la \Tr \left( \lambda \mathbf{1}_{K} - \mathbf{H} \right) \ra .
\end{equation}
Here the expectation map $\langle \ldots \rangle$ is understood to be taken w.r.t. the probability measure $P ( \mathbf{H} ) \DD \mathbf{H}$ of the random matrix. We will always have this distribution rotationally (\ie \smash{$\mathbf{H} \to \mathbf{O}^{\TT} \mathbf{H} \mathbf{O}$}, with $\mathbf{O}$ orthogonal) invariant, and hence the full information about $\mathbf{H}$ resides in its eigenvalues, distributed on average according to (\ref{eq:MeanSpectralDensityDefinition}).

On the practical side, it is more convenient to work with either of the two equivalent objects,
\begin{equation}\label{eq:GreenFunctionAndMTransformDefinition}
G_{\mathbf{H}} ( z ) \equiv \frac{1}{K} \la \Tr \frac{1}{z \mathbf{1}_{K} - \mathbf{H}} \ra , \qquad \textrm{or} \qquad M_{\mathbf{H}} ( z ) \equiv z G_{\mathbf{H}} ( z ) - 1 ,
\end{equation}
referred to as the ``Green's function'' (or the ``resolvent'') and the ``$M$--transform'' of $\mathbf{H}$. The latter is also called the ``moments' generating function,'' since if the ``moments'' \smash{$M_{\mathbf{H} , n} \equiv ( 1 / K ) \langle \Tr \mathbf{H}^{n} \rangle$} of $\mathbf{H}$ exist, it can be expanded into a power series around $z \to \infty$ as \smash{$M_{\mathbf{H}} ( z ) = \sum_{n \geq 1} M_{\mathbf{H} , n} / z^{n}$}. It should however be underlined that even for probability measures disallowing such an expansion (heavy--tailed distributions, preeminent in finance, being an example), the quantities (\ref{eq:GreenFunctionAndMTransformDefinition}) still manage to entirely capture the spectral properties of $\mathbf{H}$; hence the name ``$M$--transform'' more appropriate, in addition to being more compact.

We will show that for our purposes (multiplication of random matrices; see par.~\ref{sss:TheNTransformAndFreeRandomVariables}) the $M$--transform serves better than the Green's function. However, it is customary to write the relationship between (\ref{eq:MeanSpectralDensityDefinition}) and (\ref{eq:GreenFunctionAndMTransformDefinition}) in terms of this latter,
\begin{equation}\label{eq:SokhotskyFormula}
\rho_{\mathbf{H}} ( \lambda ) = - \frac{1}{\pi} \lim_{\epsilon \to 0^{+}} \im G_{\mathbf{H}} ( \lambda + \ii \epsilon ) = - \frac{1}{2 \pi \ii} \lim_{\epsilon \to 0^{+}} \left( G_{\mathbf{H}} ( \lambda + \ii \epsilon ) - G_{\mathbf{H}} ( \lambda - \ii \epsilon ) \right) .
\end{equation}
resulting from a well--known formula for generalized functions, \smash{$\lim_{\epsilon \to 0^{+}} 1 / ( x \pm \ii \epsilon ) = \textrm{pv} ( 1 / x ) \mp \ii \pi \delta ( x )$}.


\subsubsection{The $N$--Transform and Free Random Variables}
\label{sss:TheNTransformAndFreeRandomVariables}

The doubly correlated Wishart ensemble $\mathbf{c}$ (\ref{eq:EstimatorcDefinition}) may be viewed as a product of several random and non--random matrices. The general problem of multiplying random matrices seems formidable. In classical probability theory, it can be effectively handled in the special situation when the random terms are independent: then, the exponential map reduces it to the addition problem of independent random numbers, solved by considering the logarithm of the characteristic functions of the respective PDFs, which proves to be additive. In matrix probability theory, a crucial insight came from D.~Voiculescu and coworkers and R.~Speicher~\cite{VoiculescuDykemaNica1992,Speicher1994}, who showed how to parallel the commutative construction in the noncommutative world. It starts with the notion of ``freeness,'' which basically comprises probabilistic independence together with a lack of any directional correlation between two random matrices. This nontrivial new property happens to be the right extension of classical independence, as it allows for an efficient algorithm of multiplying free random variables (FRV), which we state below:
\begin{description}
\item[Step 1:] Suppose we have two random matrices, \smash{$\mathbf{H}_{1}$} and \smash{$\mathbf{H}_{2}$}, mutually free. Their spectral properties are best wrought into the $M$--transforms (\ref{eq:GreenFunctionAndMTransformDefinition}), \smash{$M_{\mathbf{H}_{1}} ( z )$} and \smash{$M_{\mathbf{H}_{2}} ( z )$}.
\item[Step 2:] The critical maneuver is to turn attention to the functional inverses of these $M$--transforms, the so--called ``$N$--transforms,''
    \begin{equation}\label{eq:NTransformDefinition}
    M_{\mathbf{H}} \left( N_{\mathbf{H}} ( z ) \right) = N_{\mathbf{H}} \left( M_{\mathbf{H}} ( z ) \right) = z .
    \end{equation}
\item[Step 3:] The $N$--transforms submit to a very straightforward rule upon multiplying free random matrices (the ``FRV multiplication law''),
    \begin{equation}\label{eq:FRVMultiplicationLaw}
    N_{\mathbf{H}_{1} \mathbf{H}_{2}} ( z ) = \frac{z}{1 + z} N_{\mathbf{H}_{1}} ( z ) N_{\mathbf{H}_{2}} ( z ) , \qquad \textrm{for free \smash{$\mathbf{H}_{1}$}, \smash{$\mathbf{H}_{2}$}.}
    \end{equation}
\item[Step 4:] Finally, it remains to functionally invert the resulting $N$--transform \smash{$N_{\mathbf{H}_{1} \mathbf{H}_{2}} ( z )$} to gain the $M$--transform of the product, \smash{$M_{\mathbf{H}_{1} \mathbf{H}_{2}} ( z )$}, and consequently, all the spectral properties via formula~(\ref{eq:SokhotskyFormula}).
\end{description}
It is stunning that such a simple prescription (relying on the choice of the $M$--transform as the carrier of the mean spectral information, and the construction of its functional inverse, the $N$--transform, which essentially multiplies under taking the free product) resolves the multiplication problem for free random noncommutative objects.

Let us just mention that the addition problem may be tackled along similar lines: In this case, the Green's function should be exploited, its functional inverse considered (\smash{$G_{\mathbf{H}} ( B_{\mathbf{H}} ( z ) ) = B_{\mathbf{H}} ( G_{\mathbf{H}} ( z ) ) = z$}; it is sometimes called the ``Blue's function''~\cite{Zee1996,JanikNowakPappZahed1997}), which obeys the ``FRV addition law,'' \smash{$B_{\mathbf{H}_{1} + \mathbf{H}_{2}} ( z ) = B_{\mathbf{H}_{1}} ( z ) + B_{\mathbf{H}_{2}} ( z ) - 1 / z$}, for two free random matrices. In this paper, we do not resort to using this addition formula, even though our problem could be approached through it as well.

Let us also remark that in the original mathematical formulations~\cite{VoiculescuDykemaNica1992,Speicher1994} of these frames, a slightly different language is employed: Instead of the $N$--transform, the ``$S$--transform'' is found convenient, \smash{$S_{\mathbf{H}} ( z ) \equiv ( 1 + z ) / ( z N_{\mathbf{H}} ( z ) )$}, while in place of the Blue's function, one engages the ``$R$--transform,'' \smash{$R_{\mathbf{H}} ( z ) \equiv B_{\mathbf{H}} ( z ) - 1 / z$}. They fulfil simpler laws, \smash{$S_{\mathbf{H}_{1} \mathbf{H}_{2}} ( z ) = S_{\mathbf{H}_{1}} ( z ) S_{\mathbf{H}_{2}} ( z )$} and \smash{$R_{\mathbf{H}_{1} + \mathbf{R}_{2}} ( z ) = R_{\mathbf{H}_{1}} ( z ) + Y_{\mathbf{H}_{2}} ( z )$}, respectively.


\subsubsection{Doubly Correlated Wishart Ensembles from Free Random Variables}
\label{sss:DoublyCorrelatedWishartEnsemblesFromFreeRandomVariables}

The innate potential of the FRV multiplication algorithm (\ref{eq:FRVMultiplicationLaw}) is surely revealed when inspecting the doubly correlated Wishart random matrix \smash{$\mathbf{c} = ( 1 / T ) \sqrt{\mathbf{C}} \widetilde{\mathbf{Y}} \mathbf{A} \widetilde{\mathbf{Y}}^{\TT} \sqrt{\mathbf{C}}$} (\ref{eq:EstimatorcDefinition}). This has been done in detain in~\cite{BurdaJaroszJurkiewiczNowakPappZahed2009}, so we will only accentuate the main results here, referring the reader to the original paper for a thorough explanation.

The idea is that one uses twice the cyclic property of the trace (which permits cyclic shifts in the order of the terms), and twice the FRV multiplication law (\ref{eq:FRVMultiplicationLaw}) (to break the $N$--transforms of products of matrices down to their constituents), in order to reduce the problem to solving the uncorrelated Wishart ensemble \smash{$( 1 / T ) \widetilde{\mathbf{Y}}^{\TT} \widetilde{\mathbf{Y}}$}. This last model is further simplified, again by the cyclic property and the FRV multiplication rule applied once, to the standard $\mathbf{GOE}$ random matrix squared (and the projector \smash{$\mathbf{P} \equiv \diag ( \mathbf{1}_{N} , \mathbf{0}_{T - N} )$}, designed to chip the rectangle \smash{$\widetilde{\mathbf{Y}}$} off the square $\mathbf{GOE}$), whose properties are firmly established. Let us sketch the derivation,
$$
N_{\mathbf{c}} ( z ) \stackrel{\substack{\mathrm{cyclic}\\\downarrow}}{=} N_{\frac{1}{T} \widetilde{\mathbf{Y}} \mathbf{A} \widetilde{\mathbf{Y}}^{\TT} \mathbf{C}} ( z ) \stackrel{\substack{\mathrm{FRV}\\\downarrow}}{=} \frac{z}{1 + z} N_{\frac{1}{T} \widetilde{\mathbf{Y}} \mathbf{A} \widetilde{\mathbf{Y}}^{\TT}} ( z ) N_{\mathbf{C}} ( z ) \stackrel{\substack{\mathrm{cyclic}\\\downarrow}}{=}
$$
$$
\stackrel{\substack{\mathrm{cyclic}\\\downarrow}}{=} \frac{z}{1 + z} N_{\frac{1}{T} \widetilde{\mathbf{Y}}^{\TT} \widetilde{\mathbf{Y}} \mathbf{A}} ( r z ) N_{\mathbf{C}} ( z ) \stackrel{\substack{\mathrm{FRV}\\\downarrow}}{=} \frac{z}{1 + z} \frac{r z}{1 + r z} N_{\frac{1}{T} \widetilde{\mathbf{Y}}^{\TT} \widetilde{\mathbf{Y}}} ( r z ) N_{\mathbf{A}} ( r z ) N_{\mathbf{C}} ( z ) =
$$
\begin{equation}\label{eq:DoublyCorrelatedWishartEnsembleTheMainEquation1}
= r z N_{\mathbf{A}} ( r z ) N_{\mathbf{C}} ( z ) .
\end{equation}

This is the basic formula. Since the spectral properties of $\mathbf{c}$ are given by its $M$--transform, \smash{$M \equiv M_{\mathbf{c}} ( z )$}, it is more pedagogical to recast (\ref{eq:DoublyCorrelatedWishartEnsembleTheMainEquation1}) as an equation for the unknown $M$,
\begin{equation}\label{eq:DoublyCorrelatedWishartEnsembleTheMainEquation2}
z = r M N_{\mathbf{A}} ( r M ) N_{\mathbf{C}} ( M ) .
\end{equation}
It provides a means for computing the mean spectral density of a doubly correlated Wishart random matrix once the ``true'' covariance matrices $\mathbf{C}$ and $\mathbf{A}$ are given.

In this communication, only a particular instance of this fundamental formula is applied, namely with an arbitrary auto--covariance matrix $\mathbf{A}$, but with trivial cross--covariances, \smash{$\mathbf{C} = \mathbf{1}_{N}$}. Using that \smash{$N_{\mathbf{1}_{K}} ( z ) = 1 + 1 / z$}, equation (\ref{eq:DoublyCorrelatedWishartEnsembleTheMainEquation2}) thins out to
\begin{equation}\label{eq:DoublyCorrelatedWishartEnsembleTheMainEquation3}
r M = M_{\mathbf{A}} \left( \frac{z}{r ( 1 + M )} \right) ,
\end{equation}
which will be strongly exploited below. Let us mention that these equalities (\ref{eq:DoublyCorrelatedWishartEnsembleTheMainEquation2}), (\ref{eq:DoublyCorrelatedWishartEnsembleTheMainEquation3}) have been derived through other, more tedious, techniques (the planar Feynman--diagrammatic expansion, the replica trick) in~\cite{BurdaGorlichJaroszJurkiewicz2004,BurdaJurkiewicz2004,BurdaJurkiewiczWaclaw2005-1,BurdaGorlichJurkiewiczWaclaw2006,BurdaJurkiewiczWaclaw2005-2}.


\section{VARMA from Free Random Variables}
\label{s:VARMAFromFreeRandomVariables}

In what follows, we will assume that the $\VMAq$, $\VARq$, or $\VARMA$ stochastic processes are covariance (weak) stationary; for details, we refer to~\cite{Lutkepohl2005}. It implies certain restrictions on their parameters, but we will not bother with this issue in the current work. Another consequence is that the processes display some interesting features, such as invertibility.

For all this, we must in particular take both $N$ and $T$ large from the start, with their ratio $r \equiv N / T$ fixed (\ref{eq:ThermodynamicalLimit}). More precisely, we stretch the range of the $a$--index from minus to plus infinity. This means that all the finite--size effects (appearing at the ends of the time series) are readily disregarded. In particular, there is no need to care about initial conditions for the processes, and all the recurrence relations are assumed to continue to the infinite past.


\subsection{The $\VMAq$ Process}
\label{ss:TheVMAqProcess}


\subsubsection{The Definition of $\VMAq$}
\label{sss:VMAqTheDefinition}

We consider a situation when $N$ stochastic variables evolve according to identical independent $\VMAq$ (vector moving average) processes, which we sample over a time span of $T$ moments. This is a simple generalization of the standard univariate weak--stationary moving average $\MAq$. In such a setting, the value \smash{$Y_{i a}$} of the $i$--th ($i = 1 , \ldots , N$) random variable at time moment $a$ ($a = 1 , \ldots , T$) can be expressed as
\begin{equation}\label{eq:VMAqDefinition}
Y_{i a} = \sum_{\alpha = 0}^{q} a_{\alpha} \epsilon_{i , a - \alpha} .
\end{equation}
Here all the \smash{$\epsilon_{i a}$}'s are IID standard (mean zero, variance one) Gaussian random numbers (white noise), \smash{$\langle \epsilon_{i a} \epsilon_{j b} \rangle = \delta_{i j} \delta_{a b}$}. The \smash{$a_{\alpha}$}'s are some $( q + 1 )$ real constants; importantly, they do not depend on the index $i$, which reflects the fact that the processes are identical and independent (no ``spatial'' covariances among the variables). The rank $q$ of the process is a positive integer.


\subsubsection{The Auto--Covariance Matrix}
\label{sss:VMAqTheAutoCovarianceMatrix}

In order to handle such a process (\ref{eq:VMAqDefinition}), notice that the \smash{$Y_{i a}$}'s, being linear combinations of uncorrelated Gaussian numbers, must also be Gaussian random variables, albeit displaying some correlations. Therefore, to fully characterize these variables, it is sufficient to calculate their two--point covariance function; this is straightforwardly done (see appendix~\ref{ss:TheAutoCovarianceMatrixForVMAq} for details),
\begin{equation}\label{eq:VMAqTwoPointCovarianceFunction}
\la Y_{i a} Y_{j b} \ra = \delta_{i j} A^{( 1 )}_{a b} ,
\end{equation}
where
\begin{equation}\label{eq:VMAqAutoCovarianceMatrix}
A^{( 1 )}_{a b} = \kappa^{( 1 )}_{0} \delta_{a b} + \sum_{d = 1}^{q} \kappa^{( 1 )}_{d} \left( \delta_{a , b - d} + \delta_{a , b + d} \right) , \qquad \textrm{with} \qquad \kappa^{( 1 )}_{d} \equiv \sum_{\alpha = 0}^{q - d} a_{\alpha} a_{\alpha + d} , \qquad d = 0 , 1 , \ldots , q .
\end{equation}
In other words, the cross--covariance matrix is trivial, \smash{$\mathbf{C} = \mathbf{1}_{N}$} (no correlations between different variables), while the auto--covariance matrix \smash{$\mathbf{A}^{( 1 )}$}, responsible for temporal correlations, can be called ``$( 2 q + 1 )$--diagonal.'' In the course of this article, we will use several different auto--covariance matrices, and for brevity, we decide to label them with superscripts; their definitions are all collected in appendix~\ref{ss:AListOfTheVariousAutoCovarianceMatricesUsed}.

For example, in the simplest case of $\VMAOne$, it is tri--diagonal,
\begin{equation}\label{eq:VMAOneAutoCovarianceMatrix}
A^{( 1 )}_{a b} = \left( a_{0}^{2} + a_{1}^{2} \right) \delta_{a b} + a_{0} a_{1} \left( \delta_{a , b - 1} + \delta_{a , b + 1} \right) .
\end{equation}


\subsubsection{The Fourier Transform and the $M$--Transform of the Auto--Covariance Matrix}
\label{sss:VMAqTheFourierTransformAndTheMTransformOfTheAutoCovarianceMatrix}

Such an infinite matrix (\ref{eq:VMAqAutoCovarianceMatrix}) is translationally invariant (as announced, it is one of the implications of the weak stationarity), \ie the value of any of its entries depends only on the distance between its indices, \smash{$A^{( 1 )}_{a b} = A^{( 1 )} ( a - b )$}; specifically, \smash{$A^{( 1 )} ( \pm d ) = \kappa^{( 1 )}_{d}$}, for $d = 0 , 1 , \ldots , q$, and \smash{$A^{( 1 )} ( | d | > q ) = 0$}. Hence, it is convenient to rewrite this matrix in the Fourier space,
\begin{equation}\label{eq:VMAqAutoCovarianceMatrixFourier}
\widehat{A^{( 1 )}} ( p ) \equiv \sum_{d \in \mathbb{Z}} \ee^{\ii d p} A^{( 1 )} ( d ) = \kappa^{( 1 )}_{0} + 2 \sum_{d = 1}^{q} \kappa^{( 1 )}_{d} \cos ( d p ) .
\end{equation}

In this representation, the $M$--transform of \smash{$\mathbf{A}^{( 1 )}$} is readily obtained~\cite{BurdaJaroszJurkiewiczNowakPappZahed2009},
\begin{equation}\label{eq:MomentsGeneratingFunctionForATranslationallyInvariant}
M_{\mathbf{A}^{( 1 )}} ( z ) = \frac{1}{2 \pi} \int_{- \pi}^{\pi} \dd p \frac{\widehat{A^{( 1 )}} ( p )}{z - \widehat{A^{( 1 )}} ( p )} .
\end{equation}
This integral can be evaluated by the method of residues for any value of $q$, which we do in appendix~\ref{ss:TheMTransformOfTheAutoCovarianceMatrixForVMAq}, where also we print the general result (\ref{eq:VMAqGreenFunctionOfTheAutoCovarianceMatrix}). In particular, for $q = 1$,
\begin{equation}\label{eq:VMAOneMomentsGeneratingFunctionForA}
M_{\mathbf{A}^{( 1 )}} ( z ) = \frac{z}{\sqrt{z - \left( a_{0} + a_{1} \right)^{2}} \sqrt{z - \left( a_{0} - a_{1} \right)^{2}}} - 1 ,
\end{equation}
where the square roots are principal.


\subsubsection{The Pearson Estimator of the Covariances from Free Random Variables}
\label{sss:VMAqThePearsonEstimatorOfTheCovariancesFromFreeRandomVariables}

We will be interested in investigating the spectral properties of the Pearson estimator \smash{$\mathbf{c} = ( 1 / T ) \mathbf{Y} \mathbf{Y}^{\TT} = ( 1 / T ) \widetilde{\mathbf{Y}} \mathbf{A}^{( 1 )} \widetilde{\mathbf{Y}}^{\TT}$} (\ref{eq:EstimatorcDefinition}). The $M$--transform of this correlated Wishart random matrix, \smash{$M \equiv M_{\mathbf{c}} ( z )$}, can be retrieved from equation (\ref{eq:DoublyCorrelatedWishartEnsembleTheMainEquation3}). We could write it for any $q$ using (\ref{eq:VMAqGreenFunctionOfTheAutoCovarianceMatrix}), but we will restrict ourselves only to $q = 1$, in which case the substitution of (\ref{eq:VMAOneMomentsGeneratingFunctionForA}) leads to a fourth--order polynomial (Ferrari) equation for the unknown $M$,
$$
r^{4} \left( a_{0}^{2} - a_{1}^{2} \right)^{2} M^{4} + 2 r^{3} \left( - \left( a_{0}^{2} + a_{1}^{2} \right) z + \left( a_{0}^{2} - a_{1}^{2} \right)^{2} (r + 1) \right) M^{3} +
$$
$$
+ r^{2} \left( z^{2} - 2 \left( a_{0}^{2} + a_{1}^{2} \right) (r + 2) z + \left( a_{0}^{2} - a_{1}^{2} \right)^{2} \left( r^{2} + 4 r + 1 \right) \right) M^{2} +
$$
\begin{equation}\label{eq:VMAOneMainEquation}
+ 2 r \left( z^{2} - \left( a_{0}^{2} + a_{1}^{2} \right) (2 r + 1) z + \left( a_{0}^{2} - a_{1}^{2} \right)^{2} r (r + 1) \right) M + r \left( - 2 \left( a_{0}^{2} + a_{1}^{2} \right) z + \left( a_{0}^{2} - a_{1}^{2} \right)^{2} r \right) = 0 .
\end{equation}
The FRV technique allowed us therefore to find this equation in a matter of a few lines of a simple algebraic computation. It has already been derived in~\cite{JinWangMiaoHuang2009}, and (\ref{eq:VMAOneMainEquation}) may be verified to coincide with the version given in that paper. In~\cite{JinWangMiaoHuang2009}, the pertinent equation is printed before (A.6), and to compare the two, one needs to change their variables into ours according to $y \to 1 / r$, $x \to z / r$, and \smash{$\underline{m} \to - r ( 1 + M ) / z$}. The last equality means that \smash{$\underline{m}$} and $m$ of~\cite{JinWangMiaoHuang2009} correspond in our language to the Green's functions \smash{$- r G_{\mathbf{c}} ( z )$} and \smash{$- G_{\mathbf{a}} ( z / r )$}, respectively, where \smash{$\mathbf{a} = ( 1 / N ) \mathbf{Y}^{\TT} \mathbf{Y}$} is the Pearson estimator dual to $\mathbf{c}$. As mentioned, a quick extension to the case of arbitrary $q$ is possible, however the resulting equations for $M$ will be significantly more complicated; for instance, for $q = 2$, a lengthy ninth--order polynomial equation is discovered.


\subsection{The $\VARq$ Process}
\label{ss:TheVARqProcess}


\subsubsection{The Definition of $\VARq$}
\label{sss:VARqTheDefinition}

A set--up of $N$ identical and independent $\VARq$ (vector auto--regressive) processes is somewhat akin to (\ref{eq:VMAqDefinition}), \ie we consider $N$ decoupled copies of a standard univariate $\ARq$ process,
\begin{equation}\label{eq:VARqDefinition}
Y_{i a} - \sum_{\beta = 1}^{q} b_{\beta} Y_{i , a - \beta} = a_{0} \epsilon_{i a} .
\end{equation}
It is again described by the demeaned and standardized Gaussian white noise \smash{$\epsilon_{i a}$} (which triggers the stochastic evolution), as well as $( q + 1 )$ real constants \smash{$a_{0}$}, \smash{$b_{\beta}$}, with $\beta = 1 , \ldots , q$. As announced before, the time stretches to the past infinity, so no initial condition is necessary. Although at first sight (\ref{eq:VARqDefinition}) may appear to be a more involved recurrence relation for the \smash{$Y_{i a}$}'s, it is actually easily reduced to the $\VMAq$ case: It remains to remark that if one exchanges the \smash{$Y_{i a}$}'s with the \smash{$\epsilon_{i a}$}'s, one precisely arrives at the $\VMAq$ process with the constants \smash{$a^{( 2 )}_{0} \equiv 1 / a_{0}$}, \smash{$a^{( 2 )}_{\beta} \equiv - b_{\beta} / a_{0}$}, $\beta = 1 , \ldots , q$. In other words, the auto--covariance matrix \smash{$\mathbf{A}^{( 3 )}$} of the $\VARq$ process (\ref{eq:VARqDefinition}) is simply the inverse of the auto--covariance matrix \smash{$\mathbf{A}^{( 2 )}$} of the corresponding $\VMAq$ process with the described modification of the parameters,
\begin{equation}\label{eq:VARqFromVMAq}
\mathbf{A}^{( 3 )} = \left( \mathbf{A}^{( 2 )} \right)^{- 1} .
\end{equation}
This inverse exists thanks to the weak stationarity supposition.


\subsubsection{The Fourier Transform and the $M$--Transform of the Auto--Covariance Matrix}
\label{sss:VARqTheFourierTransformAndTheMTransformOfTheAutoCovarianceMatrix}

The Fourier transform of the auto--covariance matrix \smash{$\mathbf{A}^{( 3 )}$} of $\VARq$ is therefore a (number) inverse of its counterpart for $\VMAq$ with its parameters appropriately changed,
\begin{equation}\label{eq:VARqAutoCovarianceMatrixFourier}
\widehat{A^{( 3 )}} ( p ) = \frac{1}{\widehat{A^{( 2 )}} ( p )} = \frac{1}{\kappa^{( 2 )}_{0} + 2 \sum_{d = 1}^{q} \kappa^{( 2 )}_{d} \cos ( d p )} ,
\end{equation}
where
\begin{equation}\label{eq:VARqKappas}
\kappa^{( 2 )}_{d} = \frac{1}{a_{0}^{2}} \sum_{\alpha = 0}^{q - d} b_{\alpha} b_{\alpha + d} , \qquad d = 0 , 1 , \ldots , q ,
\end{equation}
and where we define \smash{$b_{0} \equiv - 1$}.

In order to find the $M$--transform of the inverse matrix, \smash{$\mathbf{A}^{( 3 )} = ( \mathbf{A}^{( 2 )} )^{- 1}$}, one employs a general result, true for any (real symmetric) random matrix $\mathbf{H}$, and obtainable through an easy algebra,
\begin{equation}\label{eq:MHInverseFromMH}
M_{\mathbf{H}^{- 1}} ( z ) = - M_{\mathbf{H}} ( 1 / z ) - 1 .
\end{equation}
Since the quantity \smash{$M_{\mathbf{A}^{( 2 )}} ( z )$} is known for any $q$ (\ref{eq:VMAqGreenFunctionOfTheAutoCovarianceMatrix}), hence is \smash{$M_{\mathbf{A}^{( 3 )}} ( z )$} via (\ref{eq:MHInverseFromMH}), but we will not print it explicitly. Let us just give it for $q = 1$, in which case (\ref{eq:MHInverseFromMH}) and (\ref{eq:VMAOneMomentsGeneratingFunctionForA}) yield
\begin{equation}\label{eq:VAROneMomentsGeneratingFunctionForA}
M_{\mathbf{A}^{( 3 )}} ( z ) = - \frac{1}{\sqrt{1 - \frac{\left( 1 - b_{1} \right)^{2}}{a_{0}^{2}} z} \sqrt{1 - \frac{\left( 1 + b_{1} \right)^{2}}{a_{0}^{2}} z}} .
\end{equation}


\subsubsection{The Auto--Covariance Matrix}
\label{sss:VARqTheAutoCovarianceMatrix}

Despite being somewhat outside of the main line of thought of this article, an interesting question would be to search for an explicit expression for the auto--covariance matrix \smash{$\mathbf{A}^{( 3 )}$} from its Fourier transform (\ref{eq:VARqAutoCovarianceMatrixFourier}),
\begin{equation}\label{eq:VARqAutoCovarianceMatrixFromItsFourierTransform}
A^{( 3 )} ( d ) = \frac{1}{2 \pi} \int_{- \pi}^{\pi} \dd p \ee^{- \ii d p} \frac{1}{\kappa^{( 2 )}_{0} + 2 \sum_{l = 1}^{q} \kappa^{( 2 )}_{l} \cos ( l p )} ,
\end{equation}
where we exploited the fact that \smash{$\mathbf{A}^{( 3 )}$} must be translationally invariant, \smash{$A^{( 3 )}_{a b} = A^{( 3 )} ( a - b )$}. This computation would shed light on the structure of temporal correlations present in a VAR setting.

This integral is evaluated by the method of residues in a very similar manner to the one shown in appendix~\ref{ss:TheMTransformOfTheAutoCovarianceMatrixForVMAq}, and we do this in appendix~\ref{ss:TheAutoCovarianceMatrixForVARq}. We discover that the auto--covariance matrix is a sum of $q$ exponential decays,
\begin{equation}\label{eq:VARqAutoCovarianceMatrix1}
A^{( 3 )} ( d ) = \sum_{\gamma = 1}^{q} C_{\gamma} \ee^{- | d | / T_{\gamma}} ,
\end{equation}
where \smash{$C_{\gamma}$} are constants, and \smash{$T_{\gamma}$} are the characteristic times (\ref{eq:VARqAutoCovarianceMatrixCharacteristicTimes}), $\gamma = 1 , \ldots , q$; these constituents are given explicitly in (\ref{eq:VARqAutoCovarianceMatrix2}). This is a well--known fact, nevertheless we wanted to establish it again within our approach.

For example, for $q = 1$, the auto--covariance matrix of $\VAROne$ is one exponential decay,
\begin{equation}\label{eq:VAROneAutoCovarianceMatrix}
A^{( 3 )} ( d ) = \frac{a_{0}^{2}}{1 - b_{1}^{2}} b_{1}^{| d |} ,
\end{equation}
where we assumed for simplicity \smash{$0 < b_{1} < 1$} (the formula can be easily extended to all values of \smash{$b_{1}$}).


\subsubsection{The Pearson Estimator of the Covariances from Free Random Variables}
\label{sss:VARqThePearsonEstimatorOfTheCovariancesFromFreeRandomVariables}

Having found an expression for the $M$--transform of the auto--covariance matrix \smash{$\mathbf{A}^{( 3 )}$} of a $\VARq$ (\ref{eq:MHInverseFromMH}), (\ref{eq:VMAqGreenFunctionOfTheAutoCovarianceMatrix}), we may proceed to investigate the equation (\ref{eq:DoublyCorrelatedWishartEnsembleTheMainEquation3}) for the $M$--transform \smash{$M \equiv M_{\mathbf{c}} ( z )$} of the correlated Wishart random matrix \smash{$\mathbf{c} = ( 1 / T ) \mathbf{Y} \mathbf{Y}^{\TT} = ( 1 / T ) \widetilde{\mathbf{Y}} \mathbf{A}^{( 3 )} \widetilde{\mathbf{Y}}^{\TT}$} (\ref{eq:EstimatorcDefinition}). We will do this explicitly only for $q = 1$, when (\ref{eq:VAROneMomentsGeneratingFunctionForA}) leads to a fourth--order (Ferrari) polynomial equation for the unknown $M$,
$$
a_{0}^{4} r^{2} M^{4} + 2 a_{0}^{2} r \left( - \left( 1 + b_{1}^{2} \right) z + a_{0}^{2} r \right) M^{3} +
$$
\begin{equation}\label{eq:VAROneMainEquation}
+ \left( \left( 1 - b_{1}^{2} \right)^{2} z^{2} - 2 a_{0}^{2} r \left( 1 + b_{1}^{2} \right) z + \left( r^{2} - 1 \right) a_{0}^{4} \right) M^{2} - 2 a_{0}^{4} M - a_{0}^{4} = 0 .
\end{equation}
This equation has been derived by another method in~\cite{JinWangMiaoHuang2009}, and our result confirms their equation (A.8), with the change in notation, $y \to 1 / r$, $x \to z / r$, $z \to r M$.


\subsection{The \smash{$\VARMA$} Process}
\label{ss:TheVARMAq1q2Process}


\subsubsection{The Definition of \smash{$\VARMA$}}
\label{sss:VARMAq1q2TheDefinition}

The two types of processes which we elaborated on above, \smash{$\VARqOne$} and \smash{$\VMAqTwo$}, can be combined into one stochastic process called \smash{$\VARMA$},
\begin{equation}\label{eq:VARMAq1q2Definition}
Y_{i a} - \sum_{\beta = 1}^{q_{1}} b_{\beta} Y_{i , a - \beta} = \sum_{\alpha = 0}^{q_{2}} a_{\alpha} \epsilon_{i , a - \alpha} .
\end{equation}
Now it is a straightforward and well--known observation (which can be verified by a direct calculation) that the auto--covariance matrix \smash{$\mathbf{A}^{( 5 )}$} of this process is simply the product (in any order) of the auto--covariance matrices of the VAR and VMA pieces; more precisely,
\begin{equation}\label{eq:VARMAq1q2FromVARq1VMAq2}
\mathbf{A}^{( 5 )} = \left( \mathbf{A}^{( 4 )} \right)^{- 1} \mathbf{A}^{( 1 )} ,
\end{equation}
where \smash{$\mathbf{A}^{( 1 )}$} corresponds to the generic \smash{$\VMAqTwo$} model (\ref{eq:VMAqAutoCovarianceMatrix}), while \smash{$\mathbf{A}^{( 4 )}$} denotes the auto--covariance matrix of \smash{$\VMAqOne$} with a slightly different modification of the parameters compared to the previously used, namely \smash{$a^{( 4 )}_{0} \equiv 1$}, \smash{$a^{( 4 )}_{\beta} \equiv - b_{\beta}$}, for \smash{$\beta = 1 , \ldots , q_{1}$}. We have thus already made use here of the fact that the auto--covariance matrix of a VAR process is the inverse of the auto--covariance matrix of a certain corresponding VMA process (\ref{eq:VARqFromVMAq}), but the new change in parameters necessary in moving from VAR to VMA has effectively \smash{$a_{0} = 1$} w.r.t. what we had before (\ref{eq:VARqFromVMAq}); it is understandable: this ``missing'' \smash{$a_{0}$} is now included in the matrix of the other \smash{$\VMAqTwo$} process.

\begin{figure}[t]
\includegraphics[width=8.5cm]{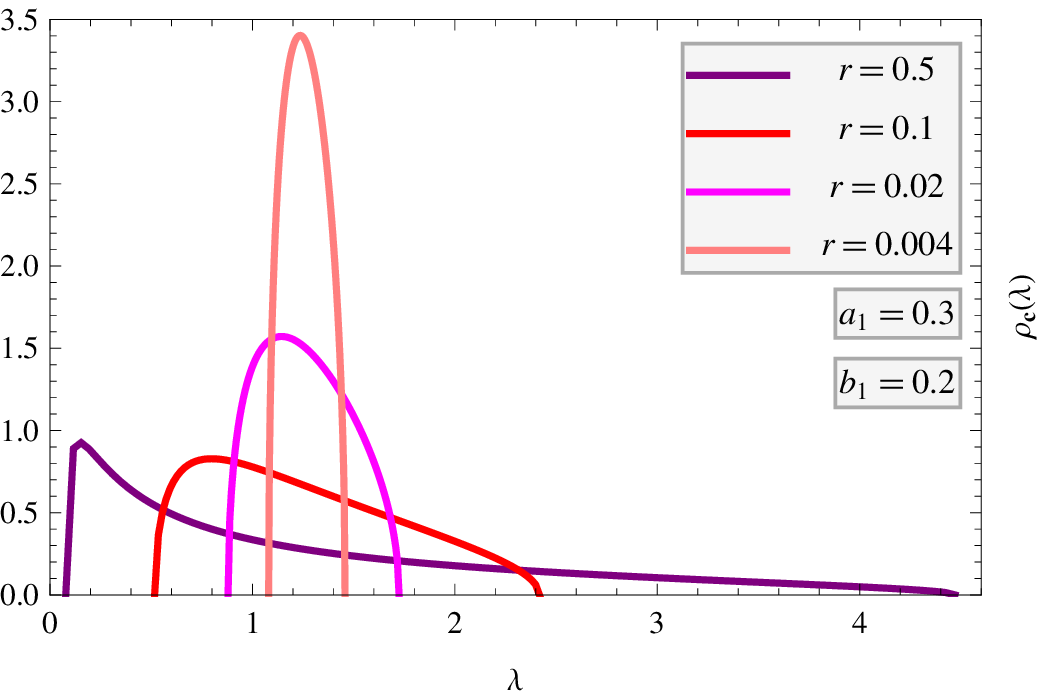}
\includegraphics[width=8.5cm]{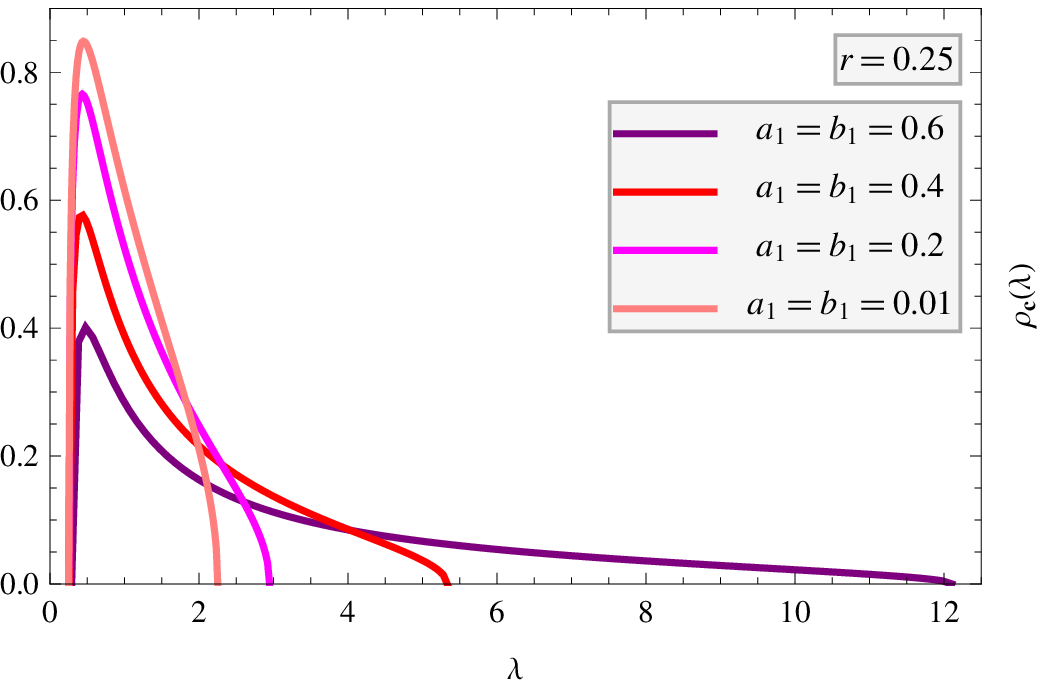}
\includegraphics[width=8.5cm]{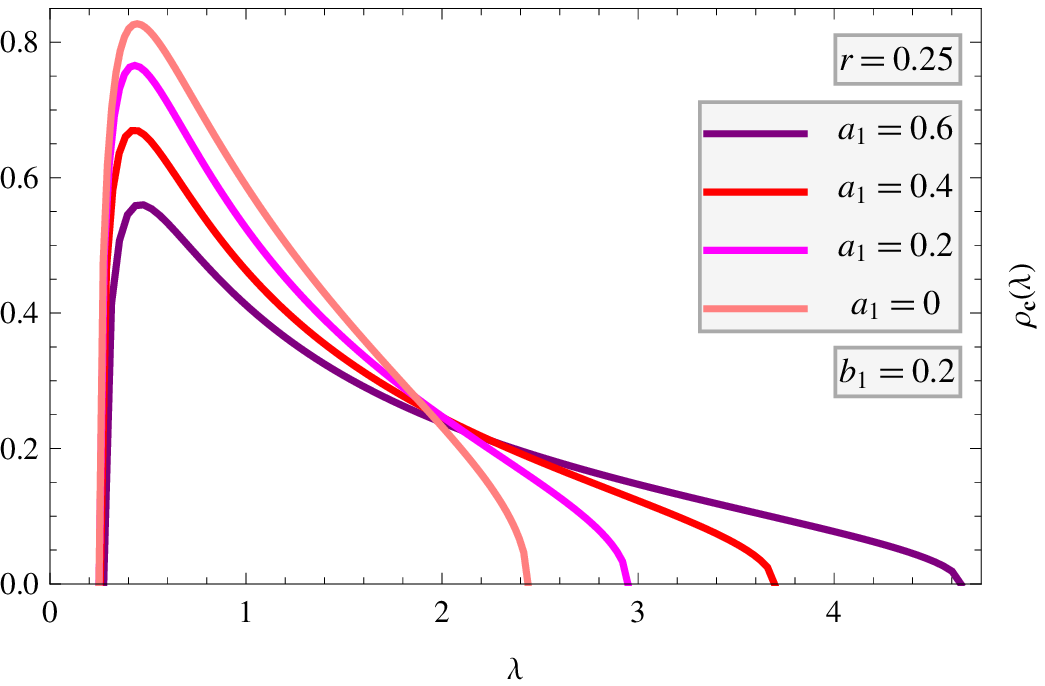}
\includegraphics[width=8.5cm]{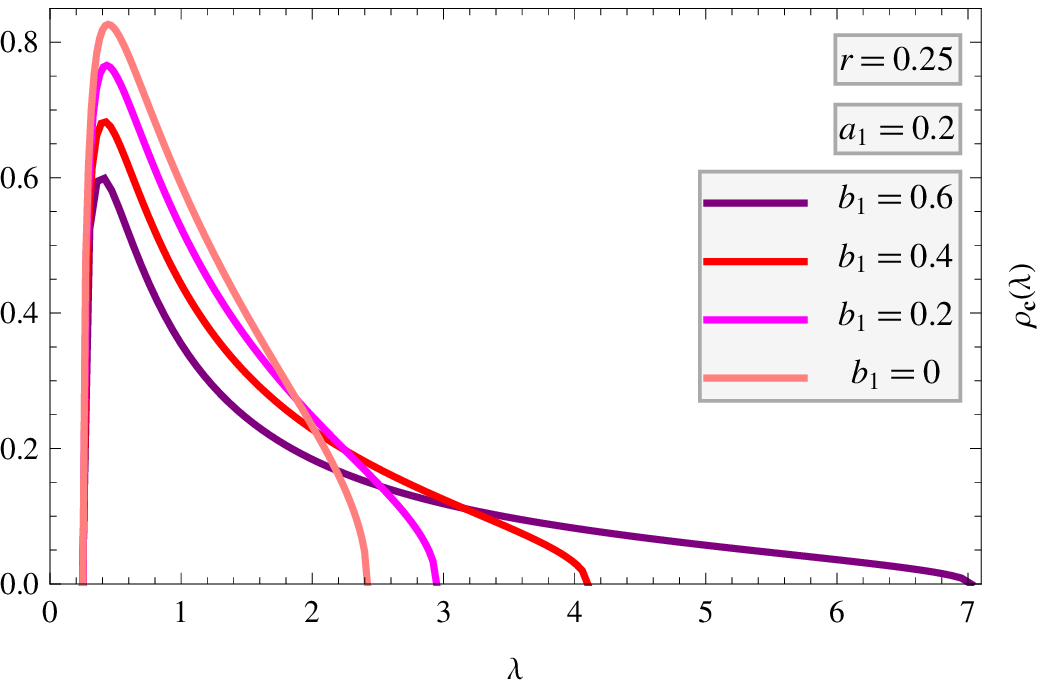}
\caption{The mean spectral density \smash{$\rho_{\mathbf{c}} ( \lambda )$} of the Pearson estimator $\mathbf{c}$ of the cross--covariances in the $\VARMAOneOne$ process computed numerically from the sixth--order polynomial equation (\ref{eq:VARMAOneOneMainEquation}), for various values of the process' parameters. The scale of these parameters is determined by choosing \smash{$a_{0} = 1$} everywhere. Recall that the theoretical formula (\ref{eq:VARMAOneOneMainEquation}) is valid in the thermodynamical limit (\ref{eq:ThermodynamicalLimit}) of $N , T \to \infty$, with $r = N / T$ kept finite.\\UP LEFT: We set the remaining VARMA parameters to \smash{$a_{1} = 0.3$}, \smash{$b_{1} = 0.2$}, while the rectangularity ratio takes the values $r = 0.5$ (the purple line), $0.1$ (red), $0.02$ (magenta), $0.004$ (pink); each one is $5$ times smaller than the preceding one. We observe how the graphs become increasingly peaked (narrower and taller) around $\lambda = 1$ as $r$ decreases, which reflects the movement of the estimator $\mathbf{c}$ toward its underlying value \smash{$\mathbf{C} = \mathbf{1}_{N}$}.\\UP RIGHT: We fix $r = 0.25$ and consider the two VARMA parameters equal to each other, with the values \smash{$a_{1} = b_{1} = 0.6$} (purple), $0.4$ (red), $0.2$ (magenta), $0.01$ (pink).\\DOWN LEFT: We hold $r = 0.25$ and \smash{$b_{1} = 0.2$}, and modify \smash{$a_{1} = 0.6$} (purple), $0.4$ (red), $0.2$ (magenta), $0.0$ (pink); for this last value, the $\VARMAOneOne$ model reduces to $\VAROne$.\\DOWN RIGHT: Similarly, but this time we assign $r = 0.25$ and \smash{$a_{1} = 0.2$}, while changing \smash{$b_{1} = 0.6$} (purple), $0.4$ (red), $0.2$ (magenta), $0.0$ (pink); this last value corresponds to $\VMAOne$.}
\label{fig:VARMAOneOneTheory}
\end{figure}

\begin{figure}[h]
\includegraphics[width=8.5cm]{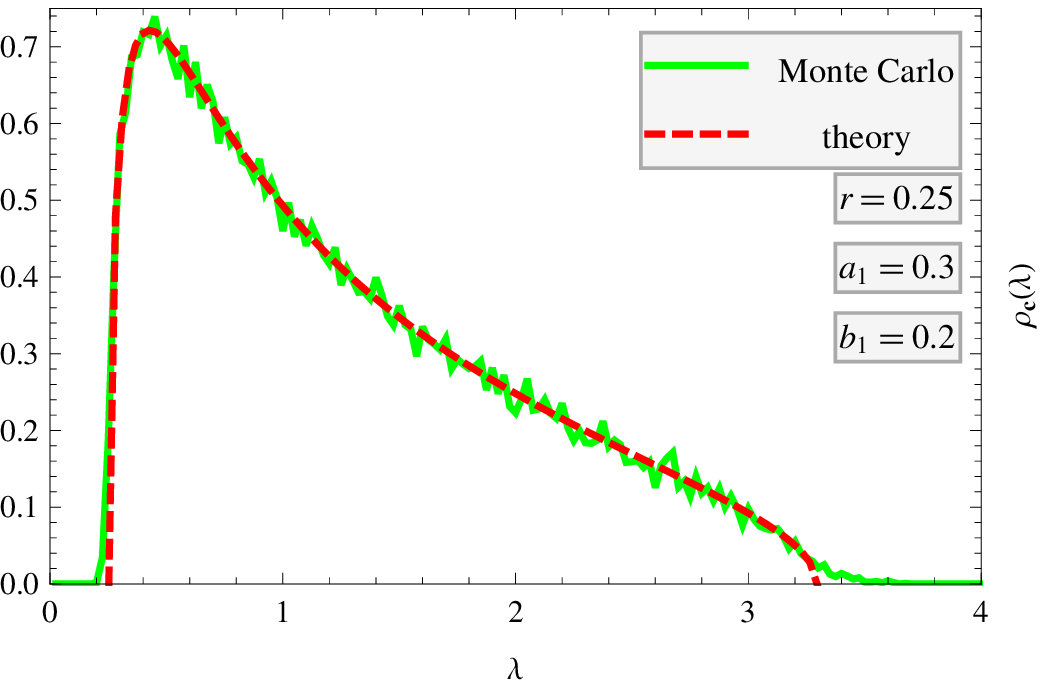}
\includegraphics[width=8.5cm]{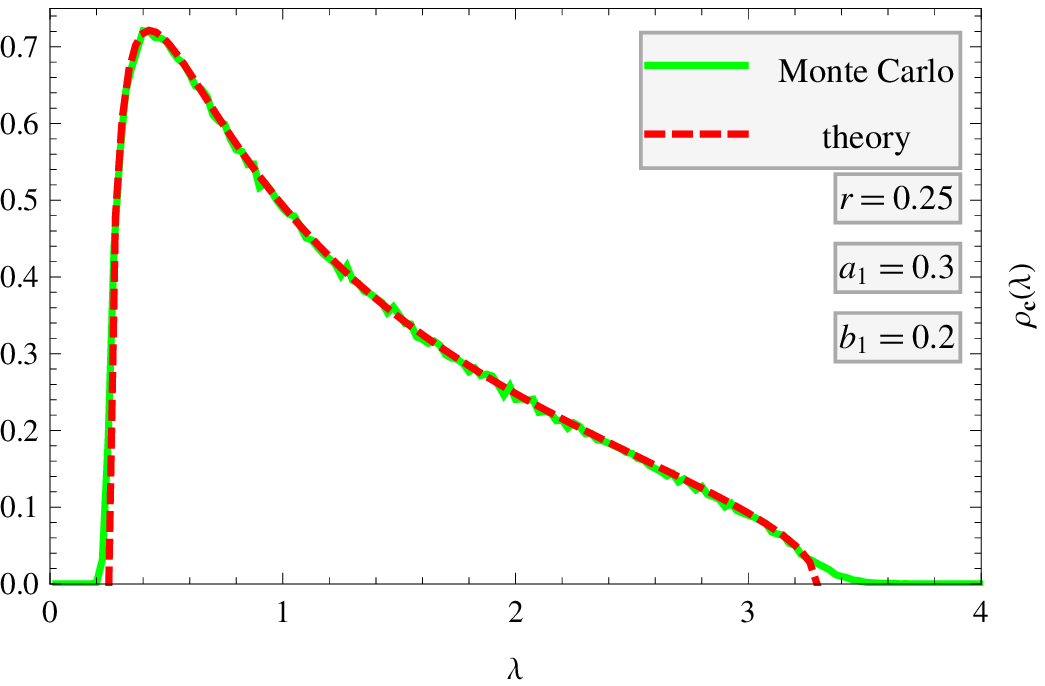}
\caption{Monte Carlo simulations of the mean spectral density \smash{$\rho_{\mathbf{c}} ( \lambda )$} (the green plots) compared to the theoretical result obtained numerically from the sixth--order equation (\ref{eq:VARMAOneOneMainEquation}) (the dashed red lines). The conformity is nearly perfect. We generate the matrices $\mathbf{Y}$ of sizes $N = 50$, $T = 200$ (\ie $r = 0.25$) from the $\VARMAOneOne$ process with the parameters \smash{$a_{0} = 1$}, \smash{$a_{1} = 0.3$}, \smash{$b_{1} = 0.2$}. The Monte Carlo simulation is repeated $1,000$ (LEFT) or $10,000$ (RIGHT) times; in this latter case, a significant improvement in the quality of the agreement is seen. One notices finite--size effects at the edges of the spectrum (``leaking out'' of eigenvalues): in the numerical simulations, $N$ and $T$ are obviously finite, while equation (\ref{eq:VARMAOneOneMainEquation}) is legitimate in the thermodynamical limit (\ref{eq:ThermodynamicalLimit}) only, hence the small discrepancies; by enlarging the chosen dimensions $50 \times 200$ one would diminish this fallout.}
\label{fig:VARMAOneOneTheoryPlusMC}
\end{figure}


\subsubsection{The Fourier Transform and the $M$--Transform of the Auto--Covariance Matrix}
\label{sss:VARMAq1q2TheFourierTransformAndTheMTransformOfTheAutoCovarianceMatrix}

The Fourier transform of the auto--covariance matrix \smash{$\mathbf{A}^{( 5 )}$} of \smash{$\VARMA$} (\ref{eq:VARMAq1q2FromVARq1VMAq2}) is simply the product of the respective Fourier transforms (\ref{eq:VMAqAutoCovarianceMatrixFourier}) and (\ref{eq:VARqAutoCovarianceMatrixFourier}),
\begin{equation}\label{eq:VARMAq1q2AutoCovarianceMatrixFourier}
\widehat{A^{( 5 )}} ( p ) = \frac{\kappa^{( 1 )}_{0} + 2 \sum_{d_{2} = 1}^{q_{2}} \kappa^{( 1 )}_{d_{2}} \cos \left( d_{2} p \right)}{\kappa^{( 4 )}_{0} + 2 \sum_{d_{1} = 1}^{q_{1}} \kappa^{( 4 )}_{d_{1}} \cos \left( d_{1} p \right)} ,
\end{equation}
where
\begin{equation}\label{eq:VARMAq1q2KappasAndLambdas}
\kappa^{( 4 )}_{d_{1}} = \sum_{\alpha_{1} = 0}^{q_{1} - d_{1}} b_{\alpha_{1}} b_{\alpha_{1} + d_{1}} , \qquad \kappa^{( 1 )}_{d_{2}} = \sum_{\alpha_{2} = 0}^{q_{2} - d_{2}} a_{\alpha_{2}} a_{\alpha_{2} + d_{2}} , \qquad d_{1} = 0 , 1 , \ldots , q_{1} , \qquad d_{2} = 0 , 1 , \ldots , q_{2} ,
\end{equation}
where we recall \smash{$b_{0} = - 1$}. For instance, for $\VARMAOneOne$ (it is described by three constants, \smash{$a_{0}$}, \smash{$a_{1}$}, \smash{$b_{1}$}), one explicitly has
\begin{equation}\label{eq:VARMAOneOneAutoCovarianceMatrixFourier}
\widehat{A^{( 5 )}} ( p ) = \frac{a_{0}^{2} + a_{1}^{2} + 2 a_{0} a_{1} \cos p}{1 + b_{1}^{2} - 2 b_{1} \cos p} .
\end{equation}

The $M$--transform of \smash{$\mathbf{A}^{( 5 )}$} can consequently be derived from the general formula (\ref{eq:MomentsGeneratingFunctionForATranslationallyInvariant}). We will evaluate here the pertinent integral only for the simplest $\VARMAOneOne$ process, even though an arbitrary case may be handled by the technique of residues,
\begin{equation}\label{eq:VARMAOneOneMomentsGeneratingFunctionForAFive}
M_{\mathbf{A}^{( 5 )}} ( z ) = \frac{1}{a_{0} a_{1} + b_{1} z} \left( - a_{0} a_{1} + \frac{z \left( a_{0} a_{1} + \left( a_{0}^{2} + a_{1}^{2} \right) b_{1} + a_{0} a_{1} b_{1}^{2} \right)}{\sqrt{\left( 1 - b_{1} \right)^{2} z - \left( a_{0} + a_{1} \right)^{2}} \sqrt{\left( 1 + b_{1} \right)^{2} z - \left( a_{0} - a_{1} \right)^{2}}} \right) .
\end{equation}


\subsubsection{The Auto--Covariance Matrix}
\label{sss:VARMAq1q2TheAutoCovarianceMatrix}

One might again attempt to track the structure of temporal covariances in a VARMA process. This can be done either by the inverse Fourier transform of (\ref{eq:VARMAq1q2AutoCovarianceMatrixFourier}), or through a direct computation based on the recurrence relation (\ref{eq:VARMAq1q2Definition}) (importantly, adhering to the assumption that it stretches to the past infinity). Let us print the result just for $\VARMAOneOne$,
\begin{equation}\label{eq:VARMAOneOneAutoCovarianceMatrix}
A^{( 5 )} ( d ) = - \frac{a_{0} a_{1}}{b_{1}} \delta_{d , 0} + \frac{\left( a_{1} + a_{0} b_{1} \right) \left( a_{0} + a_{1} b_{1} \right)}{b_{1} \left( 1 - b_{1}^{2} \right)} b_{1}^{| d |} ,
\end{equation}
where for simplicity \smash{$0 < b_{1} < 1$}. This is an exponential decay, with the characteristic time of the VAR piece, with an additional term on the diagonal.


\subsubsection{The Pearson Estimator of the Covariances from Free Random Variables}
\label{sss:VARMAq1q2ThePearsonEstimatorOfTheCovariancesFromFreeRandomVariables}

Expression (\ref{eq:VARMAOneOneMomentsGeneratingFunctionForAFive}), along with the fundamental FRV formula (\ref{eq:DoublyCorrelatedWishartEnsembleTheMainEquation3}), allow us to write the equation satisfied by the $M$--transform \smash{$M \equiv M_{\mathbf{c}} ( z )$} of the Pearson estimator \smash{$\mathbf{c} = ( 1 / T ) \mathbf{Y} \mathbf{Y}^{\TT} = ( 1 / T ) \widetilde{\mathbf{Y}} \mathbf{A}^{( 5 )} \widetilde{\mathbf{Y}}^{\TT}$} (\ref{eq:EstimatorcDefinition}) of the cross--covariances in the $\VARMAOneOne$ process; it happens to be polynomial of order six, and we print it (\ref{eq:VARMAOneOneMainEquation}) in appendix~\ref{ss:TheEquationForTheMTransformOfThePearsonEstimatorOfTheCovariancesForVARMAOneOne}. It may be solved numerically, a proper solution chosen (the one which leads to a sensible density: real, positive--definite, normalized to unity), and finally, the mean spectral density \smash{$\rho_{\mathbf{c}} ( \lambda )$} derived from (\ref{eq:SokhotskyFormula}). We show the shapes of this density for a variety of the values of the parameters $r$, \smash{$a_{0}$}, \smash{$a_{1}$}, \smash{$b_{1}$} in fig.~\ref{fig:VARMAOneOneTheory}. Moreover, in order to test the result (\ref{eq:VARMAOneOneMainEquation}), and more broadly, to further establish our FRV framework in the guise of formula (\ref{eq:DoublyCorrelatedWishartEnsembleTheMainEquation3}), the theoretical form of the density is compared to Monte Carlo simulations in fig.~\ref{fig:VARMAOneOneTheoryPlusMC}; they remain in excellent concord. These are the main findings of this article.


\section{Conclusions}
\label{s:Conclusions}

In this paper we attempted to advertise the power and flexibility of the Free Random Variables calculus for multivariate stochastic processes of the VARMA type. The FRV calculus is ideally suited for multidimensional time series problems, provided the dimensions of the underlying matrices are large. The operational procedures are simple, algebraic and transparent. The structure of the final formula which relates the moments' generating function of the population covariance and the sample covariance allows one to easily derive eigenvalue density of the sample covariance. We in detail illustrated how this procedure works for $\VARMAOneOne$, confronted the theoretical prediction with numerical data obtained by Monte Carlo simulations of the VARMA process and observed a perfect agreement.

The FRV calculus is not restricted to Gaussian variables. It also works for non--Gaussian processes, including those with heavy--tailed increments belonging to the L\'{e}vy basin of attraction, where the moments do not exist. Since the majority of data collected nowadays is naturally stored in the form of huge matrices, we believe that the FRV technique is the most natural candidate for the matrix--valued ``probability calculus'' that can provide efficient algorithms for cleaning (de--noising) large sets of data and unraveling essential but hidden correlations.


\begin{acknowledgements}
This work has been supported by the Polish Ministry of Science Grant No.~N~N202~229137 (2009--2012). AJ acknowledges the support of Clico Ltd.
\end{acknowledgements}


\appendix

\section*{Appendices}
\label{s:Appendices}


\subsection{The Auto--Covariance Matrix for $\VMAq$}
\label{ss:TheAutoCovarianceMatrixForVMAq}

In this appendix, we sketch a proof of the formula (\ref{eq:VMAqAutoCovarianceMatrix}) for the auto--covariance matrix of the $\VMAq$ process. As mentioned, since the random variables are centered Gaussian, this matrix alone suffices to completely capture all their properties. We set $i = j$; the dependence on this index may be dropped as there are no correlations here. We use the definition (\ref{eq:VMAqDefinition}) of $\VMAq$, as well as the auto--covariance structure of the white noise, \smash{$\langle \epsilon_{i a} \epsilon_{j b} \rangle = \delta_{i j} \delta_{a b}$}. This leads to
$$
A^{( 1 )}_{a b} = \la Y_{i a} Y_{i b} \ra = \sum_{\alpha = 0}^{q} \sum_{\beta = 0}^{q} a_{\alpha} a_{\beta} \la \epsilon_{i , a - \alpha} \epsilon_{i , b - \beta} \ra = \sum_{\alpha = 0}^{q} \sum_{\beta = 0}^{q} a_{\alpha} a_{\beta} \delta_{a - \alpha , b - \beta} = \ldots .
$$
The double sum is symmetrized, the index $\beta$ replaced by $d \equiv \beta - \alpha$,
$$
\ldots = \frac{1}{2} \sum_{\alpha = 0}^{q} \sum_{d = - \alpha}^{q - \alpha} a_{\alpha} a_{\alpha + d} \left( \delta_{b , a + d} + \delta_{b , a - d} \right) = \ldots ,
$$
and the order of the sums interchanged (an elegant method for this is explained in~\cite{GrahamKnuthPatashnik1994}),
$$
\ldots = \frac{1}{2} \sum_{d = - q}^{q} \left( \sum_{\alpha = \max ( 0 , - d )}^{q - \min ( 0 , d )} a_{\alpha} a_{\alpha + d} \right) \left( \delta_{b , a + d} + \delta_{b , a - d} \right) ,
$$
which, upon splitting the sum over $d$ into three pieces (from $- q$ to $- 1$, $d = 0$, and from $1$ to $q$), is quickly seen to coincide with (\ref{eq:VMAqAutoCovarianceMatrix}).


\subsection{A List of the Various Auto--Covariance Matrices Used}
\label{ss:AListOfTheVariousAutoCovarianceMatricesUsed}

For the reader's convenience, let us collect in this appendix the five auto--covariance matrices which are defined throughout this paper:
\begin{itemize}
\item By \smash{$\mathbf{A}^{( 1 )}$} we denote the auto--covariance matrix of the $\VMAq$ process with the generic constants \smash{$a_{\alpha}$}, with $\alpha = 0 , 1 , \ldots , q$, as defined in (\ref{eq:VMAqDefinition}).
\item By \smash{$\mathbf{A}^{( 2 )}$} we denote the auto--covariance matrix of the $\VMAq$ process with the constants \smash{$a^{( 2 )}_{0} \equiv 1 / a_{0}$}, \smash{$a^{( 2 )}_{\beta} \equiv - b_{\beta} / a_{0}$}, where $\beta = 1 , \ldots , q$.
\item By \smash{$\mathbf{A}^{( 3 )}$} we denote the auto--covariance matrix of the $\VARq$ process with the generic constants \smash{$a_{0}$}, \smash{$b_{\beta}$}, with $\beta = 1 , \ldots , q$, as defined in (\ref{eq:VARqDefinition}). There holds \smash{$\mathbf{A}^{( 3 )} = ( \mathbf{A}^{( 2 )} )^{- 1}$} (\ref{eq:VARqFromVMAq}).
\item By \smash{$\mathbf{A}^{( 4 )}$} we denote the auto--covariance matrix of the $\VMAqOne$ process with the constants \smash{$a^{( 4 )}_{0} \equiv 1$}, \smash{$a^{( 4 )}_{\beta} \equiv - b_{\beta}$}, where \smash{$\beta = 1 , \ldots , q_{1}$}.
\item By \smash{$\mathbf{A}^{( 5 )}$} we denote the auto--covariance matrix of the $\VARMA$ process with the generic constants \smash{$b_{\beta}$}, \smash{$a_{\alpha}$}, with \smash{$\beta = 1 , \ldots , q_{1}$} and \smash{$\alpha = 0 , 1 , \ldots , q_{2}$}, according to the definition (\ref{eq:VARMAq1q2Definition}). There is \smash{$\mathbf{A}^{( 5 )} = ( \mathbf{A}^{( 4 )} )^{- 1} \mathbf{A}^{( 1 )}$} (\ref{eq:VARMAq1q2FromVARq1VMAq2}), where in the latter piece \smash{$q = q_{2}$}.
\end{itemize}


\subsection{The $M$--Transform of the Auto--Covariance Matrix for $\VMAq$}
\label{ss:TheMTransformOfTheAutoCovarianceMatrixForVMAq}

We will derive here the $M$--transform (\ref{eq:MomentsGeneratingFunctionForATranslationallyInvariant}) of the auto--covariance matrix \smash{$\mathbf{A}^{( 1 )}$} of an arbitrary $\VMAq$ process, using the expression for its Fourier transform (\ref{eq:VMAqAutoCovarianceMatrixFourier}). It is a little simpler to consider the Green's function,
\begin{equation}\label{eq:GreensFunctionForATranslationallyInvariant}
G_{\mathbf{A}^{( 1 )}} ( z ) = \frac{1 + M_{\mathbf{A}^{( 1 )}} ( z )}{z} = \frac{1}{\pi} \int_{0}^{\pi} \dd p \frac{1}{z - \widehat{A^{( 1 )}} ( p )} ,
\end{equation}
where the integration range has been halved due to the evenness of the integrand.

This integral is performed with help of the change of variables $y \equiv 2 \cos p$. The measure, when $p \in [ 0 , \pi ]$, reads \smash{$\dd p = - \dd y / \sqrt{4 - y^{2}}$}. A basic observation is that the denominator of the integrand is a linear combination of $\cos ( d p )$, for $d = 1 , \ldots , q$, and each such a cosine can be cast as a polynomial of order $d$ in $y$ through the de Moivre formula. Hence, the denominator is a polynomial of order $q$ in $y$,
\begin{equation}\label{eq:VMAqGreenFunctionOfTheAutoCovarianceMatrixDenominator}
\widehat{A^{( 1 )}} ( p ) - z = \kappa^{( 1 )}_{0} - z + 2 \sum_{d = 1}^{q} \kappa^{( 1 )}_{d} \cos ( d p ) = \psi \prod_{\beta = 1}^{q} \left( y - y_{\beta} \right) ,
\end{equation}
where the \smash{$y_{\beta}$}'s are the $q$ roots (which we assume to be single), and $\psi$ is the coefficient at \smash{$y^{q}$}. Using the method of residues, one readily finds
\begin{equation}\label{eq:VMAqGreenFunctionOfTheAutoCovarianceMatrix}
G_{\mathbf{A}^{( 1 )}} ( z ) = - \frac{1}{\pi} \frac{1}{\psi} \int_{- 2}^{2} \dd y \frac{1}{\sqrt{4 - q^{2}}} \frac{1}{\prod_{\beta = 1}^{q} \left( y - y_{\beta} \right)} = \frac{1}{\psi} \sum_{\gamma = 1}^{q} \frac{1}{\prod_{\substack{\beta = 1\\\beta \neq \gamma}}^{q} \left( y_{\gamma} - y_{\beta} \right)} \frac{1}{\sqrt{y_{\gamma} - 2} \sqrt{y_{\gamma} + 2}} ,
\end{equation}
where the two square roots on the r.h.s. are principal. This is an explicit formula for the Green's function of \smash{$\mathbf{A}^{( 1 )}$}, provided one has factorized the order--$q$ polynomial (\ref{eq:VMAqGreenFunctionOfTheAutoCovarianceMatrixDenominator}).


\subsection{The Auto--Covariance Matrix for $\VARq$}
\label{ss:TheAutoCovarianceMatrixForVARq}

Let us argue now that the Fourier transform (\ref{eq:VARqAutoCovarianceMatrixFourier}) leads to the auto--covariance matrix of $\VARq$ (\ref{eq:VARqAutoCovarianceMatrixFromItsFourierTransform}) of the form of a sum of exponential decays (\ref{eq:VARqAutoCovarianceMatrix1}), and let us give precise expressions for the constants \smash{$C_{\gamma}$} and the characteristic times \smash{$T_{\gamma}$}, $\gamma = 1 , \ldots , q$.

We proceed by the technique of residues, analogously to appendix~\ref{ss:TheMTransformOfTheAutoCovarianceMatrixForVMAq}, however this time with aid of another variable, \smash{$x \equiv \ee^{- \ii p}$}, related to the previously used through $y = 2 \cos p = x + 1 / x$. The integration measure is $\dd p = \ii \dd x / x$, and the integration path is counterclockwise around the centered unit circle. The denominator of the integrand is a polynomial of order $q$ in the variable $y$, having thus some $q$ roots \smash{$\tilde{y}_{\beta}$}, $\beta = 1 , \ldots , q$. Therefore, there are $2 q$ corresponding solutions for the variable $x$, with a half of them inside the integration path and a half outside; let \smash{$\tilde{x}_{\beta}$} be the solutions to \smash{$x + 1 / x = \tilde{y}_{\beta}$} with the absolute values less than $1$. Only them contribute to the integral, and their residues straightforwardly give
\begin{equation}\label{eq:VARqAutoCovarianceMatrix2}
A^{( 3 )} ( d ) = \frac{1}{\psi} \sum_{\gamma = 1}^{q} \frac{\left( \tilde{x}_{\gamma} \right)^{| d | + q - 1}}{\prod_{\substack{\beta = 1\\\beta \neq \gamma}}^{q} \left( \tilde{x}_{\gamma} - \tilde{x}_{\beta} \right) \prod_{\beta = 1}^{q} \left( \tilde{x}_{\gamma} - \frac{1}{\tilde{x}_{\beta}} \right)} .
\end{equation}
This is indeed $q$ exponents \smash{$( \tilde{x}_{\gamma} )^{| d |}$}, $\gamma = 1 , \ldots , q$. Remark that the solutions may be complex, hence this is really $q$ different exponential decays \smash{$\exp ( - | d | / T_{\gamma} )$}, with the characteristic times
\begin{equation}\label{eq:VARqAutoCovarianceMatrixCharacteristicTimes}
T_{\gamma} \equiv - \frac{1}{\log \left| \tilde{x}_{\gamma} \right|}
\end{equation}
(these times are positive as the roots have the absolute values less than $1$), possibly modulated by sinusoidal oscillations when a root has an imaginary part.

For example, for $q = 1$ there is one exponential decay (\ref{eq:VAROneAutoCovarianceMatrix}), while for $q = 2$, one obtains either two exponential decays (the two roots are real and different), or one exponential decay modulated by oscillations (the two roots are complex and mutually conjugate), \emph{etc.}


\subsection{The Equation for the $M$--Transform of the Pearson Estimator of the Covariances for $\VARMAOneOne$}
\label{ss:TheEquationForTheMTransformOfThePearsonEstimatorOfTheCovariancesForVARMAOneOne}

The sixth--order polynomial equation obeyed by \smash{$M \equiv M_{\mathbf{c}} ( z )$} in the case of $\VARMAOneOne$ reads,
$$
r^{4} a_{0}^{2} a_{1}^{2} \left( a_{0}^{2} - a_{1}^{2} \right)^{2} M^{6} +
$$
$$
+ 2 r^{3} a_{0} a_{1} {\color{red}\Bigg(} {\color{blue}\bigg(} \left( a_{0}^{4} - 6 a_{0}^{2} a_{1}^{2} + a_{1}^{4} \right) b_{1} - a_{0} a_{1} \left( a_{0}^{2} + a_{1}^{2} \right) \left( b_{1}^{2} + 1 \right) {\color{blue}\bigg)} z + \left( 1 + 2 r \right) a_{0} a_{1} \left( a_{0}^{2} - a_{1}^{2} \right)^{2} {\color{red}\Bigg)} M^{5} +
$$
$$
+ r^{2} {\color{red}\Bigg(} {\color{blue}\bigg(} \left( a_{0}^{4} - 20 a_{0}^{2} a_{1}^{2} + a_{1}^{4} \right) b_{1}^{2} - 4 a_{0} a_{1} \left( a_{0}^{2} + a_{1}^{2} \right) b_{1} \left( b_{1}^{2} + 1 \right) + a_{0}^{2} a_{1}^{2} \left( b_{1}^{4} + 1 \right) {\color{blue}\bigg)} z^{2} + \Bigg.
$$
$$
+ 2 a_{0} a_{1} {\color{blue}\bigg(} \left( \left( 1 + 3 r \right) \left( a_{0}^{4} + a_{1}^{4} \right) - 2 \left( 5 + 9 r \right) a_{0}^{2} a_{1}^{2} \right) b_{1} - \left( 2 + 3 r \right) a_{0} a_{1} \left( a_{0}^{2} + a_{1}^{2} \right) \left( b_{1}^{2} + 1 \right) {\color{blue}\bigg)} z +
$$
$$
\Bigg. + \left( 1 + 8 r + 6 r^{2} \right) a_{0}^{2} a_{1}^{2} \left( a_{0}^{2} - a_{1}^{2} \right)^{2} {\color{red}\Bigg)} M^{4} +
$$
$$
+ 2 r {\color{red}\Bigg(} b_{1} {\color{blue}\bigg(} - 6 a_{0} a_{1} b_{1}^{2} - \left( a_{0}^{2} + a_{1}^{2} \right) b_{1} \left( b_{1}^{2} + 1 \right) + a_{0} a_{1} \left( b_{1}^{4} + 1 \right) {\color{blue}\bigg)} z^{3} + \Bigg.
$$
$$
+ {\color{blue}\bigg(} \left( - 10 \left( 1 + 2 r \right) a_{0}^{2} a_{1}^{2} + r \left( a_{0}^{4} + a_{1}^{4} \right) \right) b_{1}^{2} - 2 \left( 1 + 2 r \right) a_{0} a_{1} \left( a_{0}^{2} + a_{1}^{2} \right) b_{1} \left( b_{1}^{2} + 1 \right) + \left( 1 + r \right) a_{0}^{2} a_{1}^{2} \left( b_{1}^{4} + 1 \right) {\color{blue}\bigg)} z^{2} +
$$
$$
+ a_{0} a_{1} {\color{blue}\bigg(} \left( 3 r \left( 1 + r \right) \left( a_{0}^{4} + a_{1}^{4} \right) - 2 \left( 2 + 15 r + 9 r^{2} \right) a_{0}^{2} a_{1}^{2} \right) b_{1} - \left( 1 + 6 r + 3 r^{2} \right) a_{0} a_{1} \left( a_{0}^{2} + a_{1}^{2} \right) \left( b_{1}^{2} + 1 \right) {\color{blue}\bigg)} z +
$$
$$
\Bigg. + 2 r \left( 1 + 3 r + r^{2} \right) a_{0}^{2} a_{1}^{2} \left( a_{0}^{2} - a_{1}^{2} \right)^{2} {\color{red}\Bigg)} M^{3} +
$$
$$
+ {\color{red}\Bigg(} b_{1}^{2} \left( 1 - b_{1}^{2} \right)^{2} z^{4} + 2 b_{1} {\color{blue}\bigg(} -2 \left( 1 + 3 r \right) a_{0} a_{1} b_{1}^{2} - r \left( a_{0}^{2} + a_{1}^{2} \right) b_{1} \left( b_{1}^{2} + 1 \right) + \left( 1 + r \right) a_{0} a_{1} \left( b_{1}^{4} + 1 \right) {\color{blue}\bigg)} z^{3} + \Bigg.
$$
$$
+ {\color{blue}\bigg(} -\left( \left( 1 - r^{2} \right) \left( a_{0}^{4} + a_{1}^{4} \right) + 2 \left( 3 + 20 r + 10 r^{2} \right) a_{0}^{2} a_{1}^{2} \right) b_{1}^{2} - \bigg.
$$
$$
\bigg. - 2 \left( 1 + 4 r + 2 r^{2} \right) a_{0} a_{1} \left( a_{0}^{2} + a_{1}^{2} \right) b_{1} \left( b_{1}^{2} + 1 \right) + r \left( 4 + r \right) a_{0}^{2} a_{1}^{2} \left( b_{1}^{4} + 1 \right) {\color{blue}\bigg)} z^{2} +
$$
$$
+ 2 r a_{0} a_{1} {\color{blue}\bigg(} \left( r \left( 3 + r \right) \left( a_{0}^{4} + a_{1}^{4} \right) - 6 \left( 2 + 5 r + r^{2} \right) a_{0}^{2} a_{1}^{2} \right) b_{1} - \left( 3 + 6 r + r^{2} \right) a_{0} a_{1} \left( a_{0}^{2} + a_{1}^{2} \right) \left( b_{1}^{2} + 1 \right) {\color{blue}\bigg)} z +
$$
$$
\Bigg. + r^{2} \left( 6 + 8 r + r^{2} \right) a_{0}^{2} a_{1}^{2} \left( a_{0}^{2} - a_{1}^{2} \right)^{2} {\color{red}\Bigg)} M^{2} +
$$
$$
+ 2 {\color{red}\Bigg(} a_{0} a_{1} b_{1} \left( 1 - b_{1}^{2} \right)^{2} z^{3} + \Bigg.
$$
$$
+ {\color{blue}\bigg(} - \left( a_{0}^{4} + a_{1}^{4} + 2 \left( 3 + 5 r \right) a_{0}^{2} a_{1}^{2} \right) b_{1}^{2} - 2 \left( 1 + r \right) a_{0} a_{1} \left( a_{0}^{2} + a_{1}^{2} \right) b_{1} \left( b_{1}^{2} + 1 \right) + r a_{0}^{2} a_{1}^{2} \left( b_{1}^{4} + 1 \right) {\color{blue}\bigg)} z^{2} +
$$
$$
+ r a_{0} a_{1} {\color{blue}\bigg(} \left( r \left( a_{0}^{4} + a_{1}^{4} \right) - 2 \left( 6 + 5 r \right) a_{0}^{2} a_{1}^{2} \right) b_{1} - \left( 3 + 2 r \right) a_{0} a_{1} \left( a_{0}^{2} + a_{1}^{2} \right) \left( b_{1}^{2} + 1 \right) {\color{blue}\bigg)} z +
$$
$$
\Bigg. + r^{2} \left( 2 + r \right) a_{0}^{2} a_{1}^{2} \left( a_{0}^{2} - a_{1}^{2} \right)^{2} {\color{red}\Bigg)} M -
$$
$$
- b_{1} {\color{blue}\bigg(} \left( a_{0}^{4} + 6 a_{0}^{2} a_{1}^{2} + a_{1}^{4} \right) b_{1} + 2 a_{0} a_{1} \left( a_{0}^{2} + a_{1}^{2} \right) \left( b_{1}^{2} + 1 \right) {\color{blue}\bigg)} z^{2} -
$$
\begin{equation}\label{eq:VARMAOneOneMainEquation}
- 2 r a_{0}^{2} a_{1}^{2} {\color{blue}\bigg(} 4 a_{0} a_{1} b_{1} + \left( a_{0}^{2} + a_{1}^{2} \right) \left( b_{1}^{2} + 1 \right) {\color{blue}\bigg)} z + r^{2} a_{0}^{2} a_{1}^{2} \left( a_{0}^{2} - a_{1}^{2} \right)^{2} = 0 .
\end{equation}
This equation in a \texttt{Mathematica} file can be obtained from the authors upon request.


\end{document}